\documentclass[10pt,conference]{IEEEtran}
\usepackage{cite}
\usepackage{amsmath,amssymb,amsfonts}
\usepackage{graphicx}
\usepackage{enumitem}
\usepackage{textcomp}
\usepackage{xcolor}
\definecolor{linkpurple}{HTML}{A259FF}
\usepackage[hyphens]{url}
\usepackage{fancyhdr}
\usepackage{pifont}

\definecolor{vividteal}{HTML}{00BFA6}
\definecolor{amberyellow}{HTML}{FFB400}
\definecolor{warmcoral}{HTML}{FF6B6B}

\usepackage[
    colorlinks=true,
    linkcolor=blue,
    citecolor=blue,
    urlcolor=blue
]{hyperref}
\hypersetup{hypertexnames=false}
\usepackage{tcolorbox}
 \usepackage{makecell}
\usepackage{amsthm,booktabs,xspace,adjustbox,multirow}
\usepackage{algorithm}
\usepackage{algpseudocode}
\algnewcommand{\Input}[1]{\State \textbf{Input:} #1}
\algnewcommand{\Output}[1]{\State \textbf{Output:} #1}
\graphicspath{{figs/}}
\newcommand{\design}{\textsc{Chameleon}\xspace}

\newcommand{\bL}{\bar L}

\definecolor{recommendbg}{HTML}{E3DAED}

\title{Computationally Efficient Optimization of Per-Qubit Clifford Deformation for Non-uniform Biased Noise}

\def\hpcacameraready{} 

\newcommand\hpcaauthors{Won Joon Yun$^\dagger$, Andrew Nemec$^\ddagger$, and Jonathan M. Baker$^\dagger$}
\newcommand\hpcaaffiliation{The University of Texas at Austin$^\dagger$, The University of Texas at Dallas$^\ddagger$}
\newcommand\hpcaemail{Email: wonjoon.yun@utexas.edu}

\author{
    \IEEEauthorblockN{\hpcaauthors{}}
      \IEEEauthorblockA{
        \hpcaaffiliation{} \\
        \hpcaemail{}
      }
}

\begin{document}
\maketitle

\ifdefined\hpcacameraready 
  \pagestyle{empty}
\else
  \thispagestyle{plain}
  \pagestyle{plain}
\fi

\newcommand{\hpcaheight}{0mm}

\begin{abstract}
In fault-tolerant quantum computing systems with biased noise, Clifford deformation can substantially reduce the logical error rate (LER) without additional physical hardware overhead, such as extra qubits, syndrome extraction rounds, or code distance. Although Google Willow calibration data shows that 43\% of qubits exhibit strong $X/Z$ bias, existing calibration-aware deformation techniques remain impractical: (1) global searches over the $6^n$ deformation choices rely on computing-intensive simulations, and (2) local heuristics often underperform undeformed baselines.

We present \design{}, a fast, high-performance, and code-agnostic Clifford deformation compiler. We utilize our approximation to tackle a deformation problem based on an analytical bound on the LER. By minimizing this surrogate, \design{} finds an optimized deformation that empirically reduces the LER with substantially lower computational overhead. In our evaluation, using calibration models derived from real superconducting devices, \design{} demonstrates that improvements in our surrogate are strongly correlated with actual LER reductions, with an average rank correlation of $\rho=0.8$ and $\rho=0.89$--$0.94$ on the most strongly biased system. It also reduces classical computational time from $1.2$ days to $3.1$ minutes for the BB72 code. 
\textsc{Chameleon} achieves maximum LER reductions of $19\%$ ($13\%$ on average) for surface codes, $16\%$ ($7\%$) for color codes, and $10\%$ ($4\%$) for bivariate bicycle codes relative to competing baselines. The maximum gains for all code families are observed on the most strongly biased system.
\end{abstract}

\section{Introduction}\label{sec:intro}
\begin{figure}[t]
\centering
\includegraphics[width=\columnwidth]{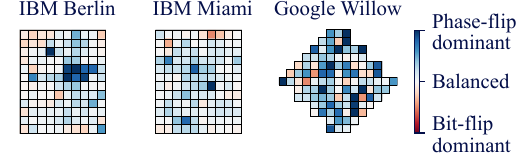}
\vspace{-15pt}
\caption{Per-qubit bit-flip/phase-flip bias, mapped onto the real qubit layout of each device. The imbalance is spatially non-uniform on every device, so no single global deformation fits it.}
\label{fig:motivation}
\vspace{-5pt}
\end{figure}

On superconducting quantum hardware, the physical noise is biased, and its strength varies from qubit to qubit across the chip~\cite{aliferis2009biased,gicev2024errorstructure}. 
Google Willow calibration data shows that $43\%$ of qubits carry a strong $X$/$Z$ imbalance whose magnitude differs by location (Figure~\ref{fig:motivation})~\cite{google2024willow}. This qubit-level bias surfaces directly at the logical level. On per-qubit noise maps derived from Willow, the logical $X$ and $Z$ failure rates of the undeformed surface code differ by up to $6.6\times$ at distance $7$.

Ideally, a logical qubit is equally protected against all types of logical errors when stored in memory. While typically logical error rates are reported as a single-valued $LER$ (Logical Error Rate), in practice this value is determined as the worst case among all error axes, for example for a single logical qubit memory as $LER$=$\max(LER_X\!,\! LER_Z)$. Each of these axes is affected by the per-qubit \textit{physical} noise and its physical bias. Fortunately, this physical bias can be controlled via \textit{Clifford deformation} where for each physical qubit, we can permute its noise vector $\mathbf{a_q}$=$(p_X, p_Y, p_Z)$ by conjugation by a single qubit Clifford operator which has no effect on the code's rate or distance.
This permutation adds no rounds, no qubits, and no code-distance cost, because it changes only the physical axis each parity check reads and never which qubits it checks. Clifford deformation can thus take a code patch whose failures are biased toward one axis, balancing the logical $X$ and $Z$ failure rates at no added cost and reducing the worst-axis LER.

There have been three major approaches. The first fixes one global deformation for a uniform, single-axis bias, as in tailored $XY$ codes~\cite{tuckett2018ultrahigh,tuckett2019tailoring,tuckett2020thresholds}, $XZZX$/$ZXXZ$~\cite{bonillaataides2021xzzx}, and bias-tailored quantum low-density parity-check (qLDPC) codes~\cite{roffe2023biastailored}. This is effective when the device bias is uniform, as in cat qubits~\cite{puri2020,guillaud2019} or dual-rail encodings~\cite{levine2024}, but today's transmon-based superconducting devices have noise that differs from qubit to qubit. In our studies across $1$,$756$ code-map instances, $29\%$ of global choices are worse than the undeformed code.

The second approach adapts to each qubit. Tiurev et al.~\cite{tiurev2023} identify the dominant axis of each qubit from the map and reassign its physical Pauli axes accordingly. This local rule is inexpensive to compute and requires no decoder. However, our studies across the same code-map instances show that the local rule underperforms the Calderbank--Shor--Steane (CSS, original) code in $27\%$ of the instances because independently suppressing each qubit's dominant error axis does not necessarily reduce the LER.

\begin{figure*}[htpb]
\centering
\includegraphics[width=\textwidth]{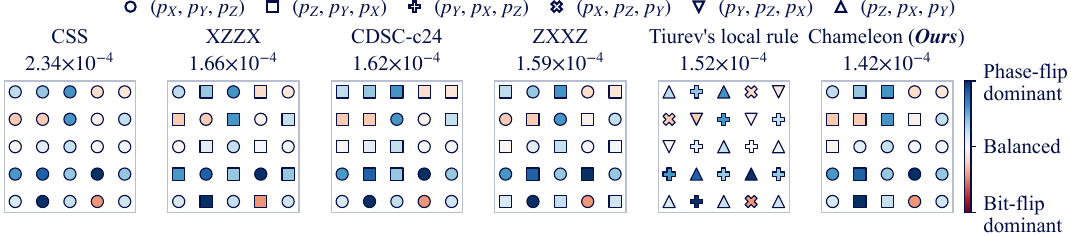}
\vspace{-10pt}
\caption{Comparison of six policies on a sampled map from Google Willow calibration data, surface code with distance 5. Values indicate LER. CDSC-c24 means CDSC validated with 24 candidates~\cite{dua2024clifford}. 
Our proposed method achieves the lowest LER without any decoder validation.}
\label{fig:overview}
\vspace{-10pt}
\end{figure*}

The third approach, Clifford-deformed surface codes (CDSC~\cite{dua2024clifford}), samples random deformation candidates, decodes each, and keeps the best, so it results in a lower LER. However, decoding to select the best is computationally infeasible because it requires Monte Carlo simulations of anywhere from $10^6$ to $10^8$ shots, and as hardware improves and failure events grow rarer, the shot count only rises. For the [[36,4,6]] bivariate bicycle (BB) code (a popular code for IBM~\cite{bravyi2024highthreshold}), there exist $10^{28}$ deformation choices. Even if only $75{,}000$ candidates are considered with an idealized 1$\mu$s-per-shot decoder, evaluating them requires about a day per calibration map ($1.2$ days for [[72,12,6]]). Thus, decoder validation inside the search loop makes deformation selection so slow that the calibration may already be stale before the deformation is determined.
Meanwhile, most of this prior work targets the surface code, and color and qLDPC codes are not well studied. Therefore, we pursue three objectives: 
(1)~eliminating decoder-in-the-loop evaluation, (2)~reducing the LER, and (3)~achieving code independence.

To achieve these goals, we must answer what precisely determines the LER and how to lower the LER via Clifford deformation.  Consider two error patterns that have the same syndrome but differ logically.  A perfect decoder takes that syndrome and selects the most likely of the two error patterns.  Logical failure occurs if the true error is not the selected one.  Thus, the LER sums the likelihood of the less likely error pattern over all syndromes. Even an optimal decoder cannot remove this \textit{ambiguity}.  It is determined by both the code and the noise. Clifford deformation does not remove these ambiguities, but it can reduce the probability mass associated with competing logical classes.

We propose \design{}, a computationally efficient LER-aware Clifford deformation approach. However, reducing the likelihood of every less likely error pattern is nontrivial:  Enumerating less likely error patterns scales exponentially with the number of qubits, and evaluating the LER is computationally intractable~\cite{iyer2015}.

To address these challenges, we adopt a two-stage approach. \textit{First,} we enumerate the error patterns that produce no syndrome but still change the logical result. We call them ambiguity operators and keep only the low-weight ones motivated by~\cite{ravi2022better}. Since the number of low-weight ambiguity operators is still large, we adopt two choices: For small codes, enumerating low-weight ambiguity operators is feasible. For larger codes, we collect the operators via search heuristics. This set depends only on the code and is independent of noise maps. Therefore, it is reused for every calibration map.

\textit{Second,} \design{} scores each deformation and finds the optimized frame (the collection of all local deformations). Evaluating the exact LER with ambiguity operators is computationally intractable because a weight-$w$ ambiguity operator contains $\mathcal{O}(2^{w})$ possible error patterns. Thus, we approximate the LER based on an analytical bound on ambiguity operators. The likelihood of logical failure in an ambiguity operator is bounded. Evaluating the upper bound is tractable because the upper bound is represented by the number of qubits involved in the ambiguity operator, which is $\mathcal{O}(w)$. Then, we model a surrogate objective by summing the upper bounds of all retained ambiguity operators and find an optimized deformation using a search algorithm.

\design{} meets the three goals: 

\textit{First,} evaluating each candidate deformation using the surrogate objective does not require decoder validation and scales linearly with the product of the number of ambiguity operators and the number of qubits. Evaluating each candidate deformation takes only microseconds for geometric codes and milliseconds for BB codes, respectively. \design{} finds an optimized deformation in 9.4 seconds, 39 seconds, and 3.1 minutes for a distance-7 surface code, a distance-7 color code, and the $[[72,12,6]]$ BB code, respectively, enabling rapid adaptation to changing calibration data.

\textit{Second,} minimizing the surrogate reduces the LER even if it is not the exact LER. The surrogate objective strongly preserves the \textit{ranking} of candidate deformations in the biased regimes where deformation is most effective. On Willow, the Spearman rank correlation between our surrogate score and the LER ranges from $\rho=0.89$ to $0.98$ across all three code families. \design{} selects the deformation with both the lowest surrogate score and the lowest measured LER among the evaluated candidates. 

\textit{Third,} \design{} supports all CSS stabilizer-code families and two fault-tolerant workloads.
Once the code-specific ambiguity operators are provided, the same surrogate scoring and optimization can be applied. 
The surrogate objective also supports worst-axis memory protection and protection of a specified logical error axis.

Figure~\ref{fig:overview} compares the undeformed CSS code with five deformation policies on a sampled Google Willow noise map for a distance-5 surface code. 
\design{} achieves the lowest LER among the evaluated policies without decoder-in-the-loop evaluation. Because \design{} modifies only the per-qubit deformation, it can be integrated into the fault-tolerant quantum computing stack as a Clifford deformation layer \textit{without additional rounds or qubits, code-distance cost, or decoder modifications}.

\newpage
\section{Background}\label{sec:bg}

\subsection{From Syndrome Ambiguity to Logical Failure}\label{sec:bg-logical-failure}

A stabilizer code repeatedly measures Pauli checks, producing a syndrome that partially reveals the error without disturbing the logical state. Each physical error is a bit-flip ($X$), a phase-flip ($Z$), or both ($Y$). For CSS codes~\cite{calderbank1996,steane1996}, $X$-type checks ($H_X$) detect $Z$-type errors and $Z$-type checks ($H_Z$) detect $X$-type errors, with logical Pauli representatives $\{\bar{L}_{X,i}, \bar{L}_{Z,i}\}^k_{i=1}$ for a code encoding $k$ logical qubits. 
The distance $d$ of a stabilizer code is defined as the minimum weight of a nontrivial logical Pauli operator and  quantifies its ability to protect encoded information. 
At low error rates, a logical failure rarely involves many qubits, since larger errors are exponentially less likely. 

A logical failure occurs when the residual error after correction is a nontrivial logical Pauli operator, thereby changing the encoded state. 
For example, two error patterns $e$ and $e'$ may produce the same syndrome while having different logical effects. 
Their product $\ell = ee'$ is a nontrivial logical Pauli operator. We call such an operator $\ell$ an \textit{ambiguity operator}. Because a syndrome cannot distinguish $e$ from $e'$, a decoder may pick the wrong one~\cite{dennis2002, gidney2021stim, harper2026failure}.

\begin{figure}[h]
\centering
\vspace{-10pt}
\includegraphics[width=\columnwidth]{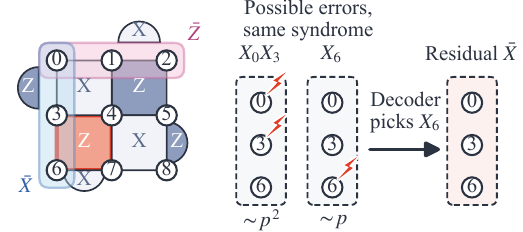}
\vspace{-20pt}
\caption{The decoding ambiguity on the distance-$3$ rotated surface code. On the left, the code carries the logical $\bar Z$ and $\bar X$, where dark plaquettes are $Z$-type checks, light plaquettes are $X$-type checks,
and orange marks triggered $Z$-type checks. On the right, the two-qubit error $X_0X_3$ (probability ${\sim}p^2$) and the single-qubit error $X_6$ (${\sim}p$) trigger exactly the same checks. The decoder picks the likelier $X_6$, and when the true error is $X_0X_3$, the residual is the logical $\bar X$. No decoder can avoid this failure; deformation can suppress it.}
\label{fig:typea-example}
\end{figure}

Under spatially uniform Pauli noise, ambiguity operators of the same weight contribute equally to the logical failure rate. Thus, the failure rate is dominated by the \textit{minimum-weight logical operators}, which determine the code distance, and by their multiplicity~\cite{fowler2013analytic,bravyi2013simulation}. 
Under the spatially non-uniform noise of real hardware, minimum-weight logical operators are not the sole factor that dictates logical failure rate, because Pauli error rates of each qubit can affect the failure rate differently.
Let $P(e)$ denote the probability of an error pattern $e$ under the calibrated noise model, and define its negative log-likelihood weight as
$\lambda(e)=-\log P(e)$, where smaller $\lambda$ means more likely. As illustrated in Figure~\ref{fig:typea-example}, we classify a failure as \textit{decoding ambiguity} (Type-A failure) when the decoder selects an error pattern $e'$ satisfying $\lambda(e')\leq \lambda(e_{\rm true})$. Because $e'$ and $e_{\rm true}$ have the same syndrome and differ by an ambiguity operator, this choice is consistent with maximum-likelihood decoding. No syndrome-based decoder can correctly resolve every such pair.

When every explanation in the class the decoder picks is less likely than the truth ($\lambda(e')>\lambda(e_{\rm true})$), a failure is due to the \textit{decoder} or \textit{algorithm} (Type-B failure). The decoder has chosen a provably worse class, so the fault is its own and a better decoder removes it. 
Of the two, we target Type-A failures. 
Their rate depends on how likely the competing error patterns are under the calibrated noise model. A code-level deformation choice can therefore change the Type-A failure rate by changing the likelihoods presented to the code. 
Type-B failures instead arise from approximations internal to the decoder and should be addressed within the decoder stack. Table~\ref{tab:typeab-bg} shows that the gap is large: exact matching~\cite{higgott2023pymatching} yields no Type-B, but approximate decoders reach ${\sim}86\%$. \textit{Thus, we use Type-A LER throughout the paper.} 

\begin{table}[htpb]
\centering
\caption{Type-A LER and Type-B share by decoder family (IBM Berlin, undeformed CSS, phenomenological, five-seed mean). Exact matching is $0\%$ Type-B, but others reach ${\sim}86\%$.}
\label{tab:typeab-bg}
\vspace{-5pt}
\renewcommand{\arraystretch}{1.0}
\setlength{\tabcolsep}{4pt}
\begin{adjustbox}{max width=\columnwidth}
\small
\begin{tabular}{|l||l|c|c|c|}
\hline
Code & Decoder & $p$ & Type-A LER & Type-B share \\
\hline\hline
surface $d{=}3$ & MWPM & 0.005 & $9.8{\times}10^{-4}$ & 0\% \\\hline
surface $d{=}5$ & MWPM & 0.005 & $1.5{\times}10^{-4}$ & 0\% \\\hline
color $d{=}3$ & Chromobius & 0.005 & $7.5{\times}10^{-4}$ & 44\% \\\hline
color $d{=}5$ & Chromobius & 0.005 & $4.9{\times}10^{-5}$ & 86\% \\\hline
BB18 & BP+OSD & 0.01 & $4.7{\times}10^{-3}$ & 42\% \\\hline
BB36 & BP+OSD & 0.01 & $9.7{\times}10^{-5}$ & 64\% \\
\hline
\end{tabular}
\end{adjustbox}
\end{table}

\subsection{Biased and Spatially Non-Uniform Noise}\label{sec:bg-noise}

Most quantum error correction (QEC) analysis assumes spatially homogeneous, Pauli-symmetric noise. However, in the superconducting hardware, processes such as relaxation, dephasing, and crosstalk produce qubit-dependent Pauli error rates $(p_{X,q},p_{Y,q},p_{Z,q})$~\cite{klimov2018,carroll2022dynamics}. 
For CSS decoding, we define the \textit{axis marginals}
$r_{X,q} = p_{X,q}+p_{Y,q}$ and
$r_{Z,q} = p_{Z,q}+p_{Y,q},$
and the per-qubit bias $\eta_q={\max(r_{X,q},r_{Z,q})}/{\min(r_{X,q},r_{Z,q})}.$
On calibration data from IBM, Zuchongzhi, and Rigetti processors, assuming identically distributed noise overestimates surface-code pseudo-thresholds by up to $95\%$~\cite{demarti2022}. This bias is strongest on Willow ($43\%$ of qubits exceed $\eta_q=1.5$), with non-trivial tails on Berlin ($13\%$) and Miami ($6\%$). 
Some hardware platforms intentionally strengthen the bias. Bias-preserving cat qubits reach second-scale bit-flip times~\cite{puri2020,guillaud2019,lescanne2020,reglade2024cat} and anchor deployed error-corrected devices~\cite{putterman2025ocelot}. These observations show that physical noise can be both strongly biased and spatially non-uniform.

\subsection{Clifford Deformation as Frame Selection}\label{sec:bg-lever}
Clifford deformation is a compile-time selection of local Pauli frames. It introduces no additional physical operation into the syndrome-extraction round. Instead, it changes how the physical $X$, $Y$, and $Z$ error rates of each qubit are presented to the code. Let $F_q$ denote the local frame assigned to physical qubit $q$. The frame of the entire code patch is written as $F=(F_1,\ldots,F_n)$, where $n$ is the number of qubits.

For decoding class $c\in\{X,Z\}$, we use $r_c^q(F)$ to denote the error rate presented by qubit $q$ under frame $F$.  
Consider a qubit $q$ assigned the Hadamard frame, such that $F_q=H$. The Hadamard frame exchanges the physical $X$ and $Z$ axes while leaving the $Y$ axis unchanged. Consequently, the two presented marginals are swapped, giving $r_X^q(F)=r_{Z,q}$ and $r_Z^q(F)=r_{X,q}$. 
By contrast, assigning $F_q=I$ leaves both presented marginals unchanged.

We first consider the \emph{binary frame space}, in which each qubit is assigned either the identity or Hadamard frame. The binary frame space is written as $\mathcal{F}_{\mathrm{binary}}=\{I,H\}^n$ and contains $2^n$ possible code-patch frames. Because the Hadamard frame exchanges only the physical $X$ and $Z$ axes, the physical $Y$ error rate continues to contribute jointly to both decoding classes throughout the binary frame space. 

We then consider the \emph{full frame space}, which allows all six rate-distinct permutations of the physical Pauli axes. We represent the code-patch frame space $\mathcal{F}_{\mathrm{full}}:=\{S_3\}^n \equiv \{I,H,S,HS,SH,HSH\}^n$ with $6^n$ possible frames. The full frame space can also change which physical Pauli axis is mapped to the code-frame $Y$ axis. The binary frame only exchanges the rates presented to the $X$ and $Z$ decoding classes. The full frame additionally changes the physical error rate assigned to the shared $Y$-error component.

\section{Limitations of Prior Work and Insights}\label{sec:limitation}
The best frame minimizes the LER on a noise map. Prior work has two drawbacks: (1) evaluating each candidate frame requires computationally intensive decoder-in-the-loop simulation, and (2) proxy objectives can diverge from the LER.

\subsection{The Decoder in the Loop Makes Computation Infeasible} \label{sec:why} 
CDSC~\cite{dua2024clifford}, a prior method that selects frames based on their decoded LER, draws \emph{random} deformations and picks the frame with the minimum LER. Finding a good frame with high probability therefore requires drawing many candidates and estimating the LER of each one. This raises two scalability problems. The first problem is that the search over the frame space grows exponentially with the number of qubits (a $6^n$ space). The second problem is decoder validation for each frame, which is a true bottleneck in classical computation. Approximating LER requires at least several million shots per candidate. The required number of shots increases as the code distance increases and as physical error shrinks. In our evaluation, validating candidates for each map requires at least a day, even after reducing the search space to 75K candidate frames. However, even after spending days to get an optimal frame, the calibration may become stale.

\begin{tcolorbox}[colback=gray!8,colframe=black!60,boxrule=0.5pt,arc=2pt,     left=6pt,right=6pt,top=6pt,bottom=6pt] \noindent \textbf{Insight 1: LER can be approximated without a decoder.} Although the exact LER requires tremendous shots, an upper bound on the exact LER can be computed analytically without any shots, reducing classical overheads.
\end{tcolorbox}
\begin{figure*}[t!]
\centering
\includegraphics[width=\textwidth]{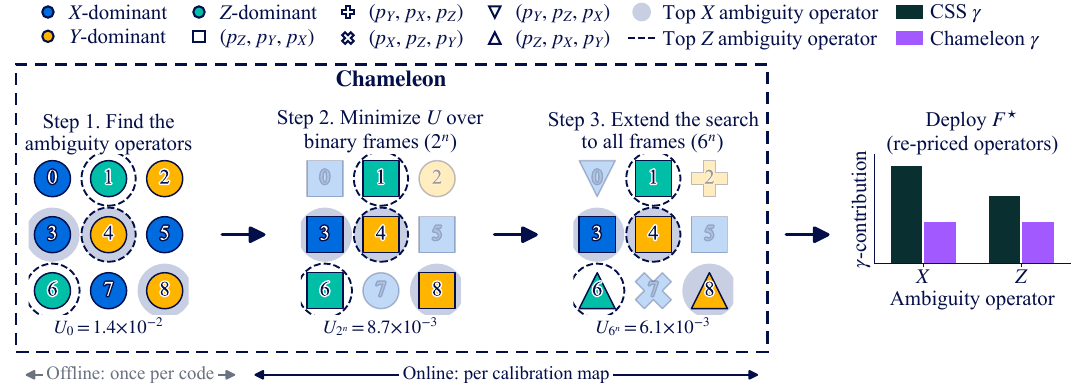}
\vspace{-15pt}
\caption{Overview of \design{}, the compile-time pass on one synthetic mixed-bias $d{=}3$ map ($\eta{=}10$, $p{=}0.005$). It takes the per-qubit noise map and the code as input. Step~1 finds the ambiguity operators of the undeformed code, and only the likeliest $X$ and $Z$ operators are drawn for legibility, a cost paid once and independent of the calibration map. Step~2 minimizes $U$, an approximation to a bound on the LER, over the binary space ($2^n$).  Step~3 extends to the full frame space ($6^n$) with the cross-entropy method, reducing the bound to $U_{6^n}$. The optimized frame $F^\star$ is then deployed without decoder validation.}

\label{fig:casestudy}
\vspace{-10pt}
\end{figure*}

\subsection{The Local Rule Cannot Track the LER}\label{sec:bg-limits}
To our knowledge, the only prior method using a per-qubit map is the local rule of Tiurev et al.~\cite{tiurev2023}.
While it considers the dominant axis of each qubit, it ignores decoding ambiguity.
We explain a concrete limitation that arises when decoding ambiguity is ignored using Figure~\ref{fig:typea-example}.
Consider the ambiguity operator $\ell{=}X_0X_3X_6$, and assume that the dominant axes of qubits 0, 3, and 6 are $Z$, $Z$, and $X$, respectively, and that the $Y$-axis is the least dominant axis for these qubits.
Even if the likelihood of the rare event $X_0X_3$ can be reduced by swapping $p_Z$ and $p_X$ for qubits 0 and 3, the local rule changes only the dominant axis of qubit 0, missing the better frame.

\begin{tcolorbox}[colback=gray!8,colframe=black!60,boxrule=0.5pt,arc=2pt,left=6pt,right=6pt,top=6pt,bottom=6pt]
\noindent \textbf{Insight 2: A good frame must consider the rare event.} 
Each ambiguity operator has a rare error pattern. Since the likelihood of the rare event dictates the LER, frame selection must reduce the likelihood of the rare event in the ambiguity operator.
\end{tcolorbox}

\section{Our Proposal: 
\design{}}\label{sec:design}

\subsection{Design Requirements and Challenges}
Our insights impose two requirements for per-qubit Clifford deformation: (1) a candidate frame should be evaluated without a decoder, and (2) the approximation of the LER must be sufficiently reliable. To satisfy the LER requirement, the naive approach based on these insights is to select the frame that minimizes the likelihood of a rare event for all ambiguity operators. This does not require any decoder validation.
However, achieving this ideal objective is not scalable. First, collecting all ambiguity operators requires a combinatorial search. Second, a single-qubit Clifford frame jointly determines the error rates presented to the logical $X$ and $Z$ classes. Consequently, a frame choice that reduces the aggregate rare-event likelihood of one logical class may increase that of the other. This solution also has computational complexity that grows exponentially with the number of qubits.

\subsection{Design Overview}
We propose \design{}, which addresses these challenges directly in three steps as follows: 

\vspace{0.05in}
\noindent \textit{Step~1. Reusable Ambiguity-Set Construction.} Because the ambiguity set depends only on the code, \design{} enumerates it \textit{once offline and reuses it across calibration maps}. \design{} keeps the \textit{low-weight} operators that dominate the LER and stores them in a lookup table (LUT).

\vspace{0.05in}
\noindent \textit{Step~2. Binary-Frame Search with an Analytic Bound.} For each ambiguity operator in the LUT, \design{} replaces the exact rare-event likelihood with an upper bound. This bound decomposes into a product of per-qubit terms. Consequently, an optimized binary frame (considering only $X$ and $Z$) can be obtained in reasonable time by minimizing the bounds for all ambiguity operators.

\vspace{0.05in}
\noindent \textit{Step~3. Refinement to Full Frame.}
Step~2 leaves the $Y$-dominant qubits unoptimized. Therefore, \design{} uses the cross-entropy method (CEM), warm-started with the optimized binary frame. This enables \design{} to select an optimized frame from the full frame space.

\newpage

We explain the pipeline of \design using a $d{=}3$ surface code in Figure~\ref{fig:casestudy}.
Each number indicates a qubit index. In \textit{Step~1}, \design takes code information as input and enumerates all low-weight ambiguity operators. Here, we indicate only the most likely ambiguities of the logical classes $X$ and $Z$. In \textit{Step~2}, \design finds the optimized frame over the binary frame space $\mathcal{F}_{\rm binary}$ with the minimum surrogate $U_{2^n}$, which approximates the sum of the likelihoods of the rare events.
For example, the top $X$ ambiguity operator crosses qubits $3$, $4$, and $8$, and the possible error patterns are $X_3X_4$ versus $X_8$. To minimize the likelihood of rare event, \design makes qubit 3 become Z-dominant. However, qubit 4 is a $Y$-dominant qubit, so it cannot be changed since Step~2 only considers $X$ or $Z$. This same process is also applied to the top $Z$ ambiguity operator.
In \textit{Step~3}, \design expands the frame space from $2^n$ to $6^n$, starting from the optimized frame in the binary frame space. Using the cross-entropy method, it changes each axis to $Y$ based on a top candidate and then discards it. This process continues until the best candidate is obtained. As a result, the $Y$-dominant center qubit $4$ can only trade its two rare axes, since a $Y$-heavy rate burdens one decoding axis wherever the frame places it.
By doing this, the surrogate cost decreases on both axes, which also leads to a reduction in the LER. After \design obtains the optimized frame over the full frame space $6^n$, it deploys the frame.

\subsection{Enumerating an Ambiguity Set}\label{sec:mechs}

A syndrome cannot uniquely determine the exact physical error. Instead, many error patterns can produce the same syndrome. These patterns should be considered when constructing the ambiguity set $\mathcal{A}$. However, the number of such patterns grows exponentially with the code size, which makes enumeration, and storing in LUT impractical. Table~\ref{tab:ambset} shows that the complete class-$X$ ambiguity set ($\mathcal{A}_X$) contains GiB-scale storage for surface $d{=}7$ and grows to PiB scale for BB72.

\begin{table}[htpb]
\centering
\vspace{0pt}
\caption{Reduction of the complete ambiguity space to the retained operator set and LUT size}
\label{tab:ambset}
\renewcommand{\arraystretch}{1.15}
\setlength{\tabcolsep}{4pt}
\small
\begin{tabular}{|l||c|c||c|c|}
\hline
\multirow{2}{*}{\makecell{~~~~Code}} & \multicolumn{2}{c||}{Class-$X$ operators} & \multicolumn{2}{c|}{LUT Size} \\
\cline{2-5}
 & \makecell{Complete\\space} & \makecell{Retained\\(\textit{Ours})} & \makecell{Complete\\space} & \makecell{Retained\\(\textit{Ours})} \\
\hline\hline
Surface $d{=}7$ & $1.68{\times}10^{7}$ & $\mathbf{2{,}497}$ & $12.3$\,GiB & $\mathbf{1.9}$\,\textbf{MiB} \\
\hline
Color $d{=}7$ & $2.62{\times}10^{5}$ & $\mathbf{4{,}078}$  & $148$\,MiB & $\mathbf{2.3}$\,\textbf{MiB} \\
\hline
BB72 & $4.40{\times}10^{12}$ & $\mathbf{45{,}483}$ & $4.5$\,PiB & $\mathbf{50}$\,\textbf{MiB} \\
\hline
\end{tabular}

\end{table}

\design{} stores only the low-weight operators. At low physical error rates, the likelihood of an error event decreases exponentially with the number of qubits involved. Even if the number of high-weight ambiguity operators is large, each of their contributions to the LER is marginal. However, the low-weight subset of ambiguity operators contributes directly to the LER. Therefore, \design{} stores all ambiguity operators up to a code-specific weight cutoff $W$ in the LUT.

Now we explain how \design{} constructs the retained set depending on the code. For the surface and color codes and for BB18 and BB36, the enumeration of low-weight subsets of ambiguity operators remains tractable as a \textit{one-time offline computation}.
For these codes, \design{} retains every ambiguity operator of weight at most $W$.
However, for the BB72 code or larger qLDPC codes, enumerating the low-weight ambiguity set is infeasible. \design{} utilizes randomized Gaussian elimination~\cite{pryadko2022qdistrnd} to reduce one representative of each nontrivial logical combination to low weight. The randomized Gaussian elimination is repeated to reduce the number of high-weight ambiguity operators. Then, \design{} expands it by the translation orbit of BB codes.
By default, \design{} uses the cutoff $W=d+2$ for surface and color codes and small BB codes (BB18 and BB36). For larger BB codes, we use the cutoff $W=d+10$ and set the number of iterations for the BB72 search to $800$.

\subsection{An Analytic Bound}\label{sec:impact-surrogate}\label{sec:obj} 
After constructing the ambiguity set from Step~1, \design{} must evaluate the rare-event likelihood associated with each operator under a candidate frame. 
We explain this with an ambiguity operator $\ell=X_0X_3X_6$. On its three qubits, there are eight possible error patterns, represented as $000,\ldots,111$. Applying $\ell$ maps each pattern to another pattern with the same syndrome but a different logical outcome. The eight patterns therefore form four competing pairs: $\{I,X_0X_3X_6\}$, $\{X_0,X_3X_6\}$, $\{X_3,X_0X_6\}$, and $\{X_6,X_0X_3\}$. The exact rare-event likelihood of $\ell$ compares the two pattern probabilities in every pair and sums the smaller probability. In general, a weight-$w$ ambiguity operator has $2^w$ possible patterns on the $w$ qubits involved in the operator and $2^{w-1}$ such pairs. Thus, exactly evaluating its rare-event likelihood requires time exponential in weight $w$. 

\design{} avoids explicitly enumerating the $2^{w-1}$ competing pairs of each ambiguity operator by using the Bhattacharyya bound~\cite{bhattacharyya1943measure}. Under independent noise, their aggregate rare-event likelihood is \textit{upper-bounded}, up to a frame-independent constant, by multiplying one Bhattacharyya coefficient per qubit (see Appendix for the full derivation):
\begin{equation} 
\label{eq:operator-bound} 
\textstyle{\Gamma_\ell(F) = \prod_{q\in\ell}\gamma\bigl(r_c^q(F)\bigr), \qquad \gamma(r)=2\sqrt{r(1-r)}.} 
\end{equation} 
Here, $r_c^q(F)$ is the class-$c$ error rate presented by qubit $q$ under frame $F$ in any frame space. A value of $\gamma$ near one indicates an error-prone presented axis, whereas a value near zero indicates a rare axis~\cite{sason2006performance}. Thus, $\Gamma_\ell(F)$ evaluates an operator using one term per qubit instead of comparing all $2^{w-1}$ competing pairs. The per-operator evaluation cost is reduced from $\mathcal{O}\!(\!2^w\!)$ to $\mathcal{O}\!(\!w\!)$.

 A logical failure in class $c$ can arise from any retained ambiguity operator in $\mathcal{A}_c$. Therefore, \design{} aggregates their individual bounds into a class-specific score: 
 \begin{equation} \label{eq:class-score} \textstyle{U_c(F) = \sum_{\ell\in\mathcal{A}_c}\Gamma_\ell(F), \qquad c\in\{X,Z\}. }
 \end{equation} 
 Each retained operator is counted separately, so operators that involve different physical qubits contribute according to the error rates presented at their respective locations. Consequently, $U_c(F)$ upper-bounds the aggregate rare-event mass of the retained ambiguity operators in logical class $c$.
The appropriate aggregation across logical classes depends on the application.

\vspace{0.05in}
\noindent \textbf{Two-axis optimization.} A QEC memory must protect an arbitrary logical state, so its reliability is limited by the worse of the logical $X$ and $Z$ classes. \design{} therefore selects \begin{equation} \label{eq:memory-objective} \textstyle{F_{\mathrm{mem}}^\star = \arg\min_F \max\left\{U_X(F),U_Z(F)\right\}.} \end{equation} This objective prevents the frame search from improving one logical class by severely degrading the other.  

\vspace{0.05in}
\noindent \textbf{One-axis optimization.} 
In contrast to a full QEC memory, some workloads depend on only one designated logical class $c^\star$. For a single-axis workload such as magic-state preparation, \design{} minimizes the corresponding logical class score: \begin{equation} 
\label{eq:single-axis-objective} \textstyle{F_{\mathrm{cons}}^\star = \arg\min_F U_{c^\star}(F).}
\end{equation} 

Thus, the same decoder-free class scores support both worst-class protection for QEC memories and targeted protection for single-axis workloads.

The weighted-distance view makes Insight~2 operational. For each qubit $q$, define its class-$c$ cost under frame $F$ as $w_c^q(F)=-\log\gamma(r_c^q(F))$. For an ambiguity operator $\ell$, its weighted cost is the sum of the costs of the qubits involved in that operator: 
\begin{equation} 
\label{eq:weighted-distance} \textstyle{d_\ell^c(F) = \sum_{q\in\ell} w_c^q(F) = -\log\Gamma_\ell(F).}
\end{equation}
A frame increases $d_\ell^c(F)$ by presenting rarer class-$c$ error rates on the qubits involved in $\ell$, thereby reducing the operator score $\gamma_\ell(F)$. The desired frame therefore cannot be obtained by independently suppressing the dominant error axis of each qubit. Instead, it must coordinate the local frame choices across the qubits of the lowest-cost ambiguity operators, which dominate $U_c(F)$ at low physical error rates. Minimizing $U_c(F)$ thus raises the effective weighted distance of the most vulnerable same-syndrome logical confusions rather than treating all qubits as equally reliable.

\subsection{Search over Frame Space}\label{sec:search} 

We introduce the search algorithm used in \design{}.

\vspace{0.05in}
\noindent\textbf{Binary-frame search (Step 2).} The objectives above specify how to score a candidate frame, but exhaustively evaluating all $2^n$ binary frames remains infeasible. \design{} therefore applies the cross-entropy method (CEM~\cite{de2005tutorial}) over the binary frame space. The search maintains one Bernoulli parameter per qubit, where each sampled bit selects either $I$ or $H$ frame. Because these frames exchange only the physical $X$ and $Z$ axes, the physical $Y$ rate remains assigned to the shared $Y$-error component throughout the binary search. By default, \design{} runs $15$ iterations with $200$ candidate frames per iteration and retains the lowest-scoring $10\%$ as the elite set. The Bernoulli parameters are initialized to $0.5$, updated toward the elite population, and clipped to $[0.02,0.98]$ to avoid premature convergence. The search evaluates $3{,}000$ candidates and returns the lowest-scoring binary frame as the warm start for the full frame search.

\vspace{0.05in}
\noindent\textbf{Full frame search (Step 3).} The binary search exchanges only the physical $X$ and $Z$ axes, leaving the physical $Y$ rate in the shared $Y$-error component.
\design{} therefore expands the search space to all six frames $\mathcal{F}_{\rm full}$. 
It applies the same CEM procedure using a per-qubit categorical  distribution over the six frames and evaluates candidates with  the same application-specific objective. The binary frame from the previous stage is used to warm-start this search, and the  lowest-scoring frame found becomes the full-frame candidate. The optimized binary frame initializes the full-frame search, but the resulting full-frame candidate is deployed only if it improves the objective by more than $\tau$ relative to the binary frame. This selection rule accepts the reassignment of the physical $Y$ rate to another axis only when it provides a benefit. We set $\tau=20\%$ by default.

\vspace{0.05in}
We summarize our design in Algorithm~\ref{alg:cham} (See Appendix).

\section{Evaluation Methodology}\label{sec:setup}
\subsection{Figures of Merit}\label{sec:figure_of_merit}
    We evaluate \design{} by addressing three questions. First, does the decoder-free surrogate preserve the LER ranking of candidate frames? Second, can \design{} select a frame within a practical compilation time budget after receiving a calibration map? Third, does the selected frame reduce the application-relevant LER  across different codes and noise maps? We answer these questions using rank correlation, frame-selection latency, and LER, respectively. 

  Spearman's coefficient~\cite{moore1996basic} measures whether the surrogate and LER rank candidate frames in the same order. Frame-selection latency measures the compilation time from receiving a  calibration map to returning a deployable frame, excluding the one-time ambiguity-set construction. 
  For a QEC memory, we report the worst-axis LER,  $\max\{\mathrm{LER}_X,\mathrm{LER}_Z\}$; for a one-axis workload, we report the LER of class  $Z$. \textit{Higher rank correlation and lower latency and LER are preferred.}

\subsection{Codes and Decoders}
We consider two \emph{geometric codes} (the rotated surface code and the $6.6.6$ planar color code at $d{=}3,5,7$) and bivariate-bicycle (BB) codes (BB18 $[[18,4,4]]$, BB36 $[[36,4,6]]$~\cite{wang2024coprime}, and BB72 $[[72,12,6]]$~\cite{bravyi2024highthreshold}).
To obtain precise results, we consider code-specialized decoders. The surface code uses MWPM~\cite{higgott2022pymatching}, the color code uses approximate Chromobius~\cite{gidney2023chromobius}, and the BB codes use iterative BP+OSD~\cite{roffe2020bposd}. Each decoder returns a logical failure if a failure exists on any of the corresponding code's logical qubits.

\subsection{Noise Model}
\noindent \textbf{Phenomenological Model.} 
Following other Clifford deformation work~\cite{dua2024clifford,das2026cliffordldpc,tiurev2023,xu2023tailoredxzzx}, we consider the phenomenological noise model~\cite{bonillaataides2021xzzx,tuckett2019tailoring,tuckett2020thresholds} due to the length of time needed for the circuit-level simulation.

\vspace{0.05in}
\noindent \textbf{Calibration-Derived Noise Maps.}
Every field is a distribution over full three-class channels $(p_X,p_Y,p_Z)$, one per qubit. The measured fields Berlin, Miami, and Willow resample per-qubit triples from Qiskit fake-backends~\cite{qiskit2024} and Willow calibrations~\cite{google2024willow} (measurable at scale~\cite{flammia2020pauli,vandenberg2023probabilistic}), twirled and rescaled to $p$. 
For geometric codes with $d\leq 5$, we use $p=0.005$; for distance-7 geometric codes and the BB-code family, we use $p=0.01$.

\vspace{0.05in}
\noindent\textbf{Synthetic Noise Maps.}
The calibration maps carry no $Y$-dominant qubits, so bias strength and $Y$-dominant mass cannot vary independently, and measured transmon fields alone cannot exercise the full frame space. Thus, we consider synthetic xyz field where each qubit's total rate fixed and draws its dominant axis uniformly over $\{X,Y,Z\}$ at bias $\eta$. By default, we use the modest biased level $\eta{=}10$ for a fair comparison.

\subsection{Baselines}\label{sec:baselines}
We compare against the undeformed CSS code, the fixed-pattern family, the local rule proposed by Tiurev et al.~\cite{tiurev2023} and CDSC. The fixed patterns are the checkerboards $XZZX$ and $ZXXZ$~\cite{bonillaataides2021xzzx}.
To our knowledge the only prior work utilizing a per-qubit map is the local rule proposed by Tiurev et al.~\cite{tiurev2023}. CDSC~\cite{dua2024clifford} is the state-of-the-art \textit{random-deformation} strategy, instantiated with a $24$-candidate decoder-validation budget for the main result table. Additionally, we increase the decoder budget of CDSC with up to $10^5$ candidates to evaluate the search effectiveness at a higher rate $p=0.02$.
Reported gains are computed relative to the best prior baseline.

\subsection{Experimental Setup}
To reduce the variance of policy comparisons, all policies use common random numbers generated from shared seeds. To estimate LER well, we collect at least 500 events for every result.  
All the experiments run on stim~\cite{gidney2021stim}, PyMatching~\cite{higgott2022pymatching}, Chromobius~\cite{gidney2023chromobius}, and ldpc~\cite{roffe2020bposd}. All baseline software implementations are used exactly as provided in \cite{qecsim, tiurev_github}.

\design{} is open source at \href{https://github.com/WonJoon-Yun/Chameleon.git}{GitHub Repository}~\cite{chameleon_github}.

\begin{figure*}[!t]\centering\includegraphics[width=\textwidth]{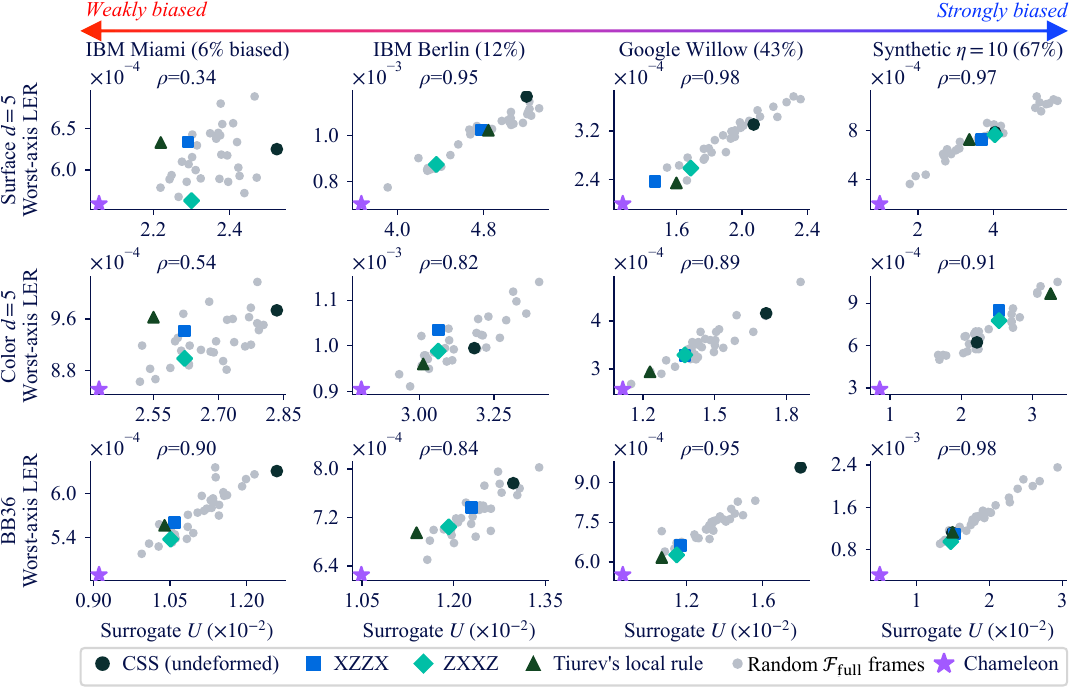}

\vspace{-5pt} 
\caption{Surrogate $U$ vs.\ measured worst-axis LER for 35 frames per cell (4 fixed baselines, the deployed \design{} frame, and 30 random full frames), at code capacity with $p{=}0.02$ and $\geq$256 ambiguity events per frame. Columns are ordered by the bias strength of the map (fraction of qubits with $\eta_q{>}1.5$, from left to right $6$, $13$, $43$, $67\%$). 
Surrogate fidelity rises with bias. 
On the least-biased map (Miami) the surface-code frames are a near-flat tie and ranking is noise-dominated, while on BB36 $\rho$ reaches $0.84$--$0.98$ on every map, and \design{} lands in the lowest-$U$, lowest-LER corner of every panel.}\label{fig:typea-scatter}
\vspace{-10pt} 
\end{figure*}

\section{Results}\label{sec:eval}

\subsection{Surrogate Fidelity}\label{sec:res-fidelity}
We measure how faithfully the surrogate $U$ orders frames according to the worst-axis LER. Figure~\ref{fig:typea-scatter} shows that the rank correlation $\rho$ increases with the bias strength of the map. On a less biased device, IBM Miami, it exhibits weak rank correlation ($\rho{=}0.34$ for surface code and $0.54$ for color code). However, on a device with strongly biased qubits (Google Willow), the rank correlation increases to $\rho{=}0.89$–$0.98$. For the delocalized qLDPC code (BB36), the surrogate is already faithful on the least-biased Miami map ($\rho{=}0.90$) and reaches $0.84$–$0.98$ across all maps.
Across the code families, lower surrogate values are empirically associated with lower LER under strongly biased noise.

\subsection{Classical Overhead Reduction}\label{sec:res-overhead}
Next, we investigate how \design reduces the classical overhead of obtaining a frame from a calibration map. We compare \design with the decoder-validated method CDSC with $75{,}000$ candidates. \design{} takes 3.1 minutes on BB72 and under 40 seconds on the geometric codes. The baseline column uses an idealized $1\,\mu$s-per-shot decoder favorable to the baseline, and the measured BP+OSD decoder is about $3.3{\times}10^{3}$ times slower per shot. 
Per map, \design{} replaces all $75{,}000$ candidate decodes of about $10^6$ shots each with $75{,}000$ deterministic surrogate evaluations, and the one-time enumeration of the ambiguity set is amortized after the first map.

\begin{table}[htpb]
\centering
\caption{Pricing all $75{,}000$ candidates by decoding takes days per map, even with an idealized $1\,\mu$s/shot decoder, while the surrogate-CEM search takes seconds to minutes.}

\label{tab:decodecost}


\renewcommand{\arraystretch}{1.2}

\setlength{\tabcolsep}{1pt}

\begin{adjustbox}{max width=\columnwidth}

\small

\begin{tabular}{|l||c|c|c|c|c|}

\hline

Code & LER & Shots/cand. & $1$ cand. & \shortstack{Search\\(all cand.)} & \shortstack{Search\\\textit{\textbf{(Ours)}}} \\

\hline\hline

Surface $d{=}7$ & $1.2\times10^{-5}$ & $8.3{\times}10^{6}$ & 8.3 sec & \textcolor{black}{7.2 \textit{days}} & \textcolor{black}{9.4 \textit{sec}} \\

Color $d{=}7$ & $2.1\times10^{-5}$ & $4.8{\times}10^{6}$ & 4.8 sec & \textcolor{black}{4.2 \textit{days}} & \textcolor{black}{39 \textit{sec}} \\

BB72 & $7.3\times10^{-5}$ & $1.4{\times}10^{6}$ & 1.4 sec & \textcolor{black}{1.2~\textit{days}} & \textcolor{black}{3.1 \textit{min}} \\

\hline
\end{tabular}
\end{adjustbox}
\end{table}

\subsection{Search Effectiveness}

\begin{figure}[htpb]
\centering
\vspace{-5pt}

\includegraphics[width=\columnwidth]{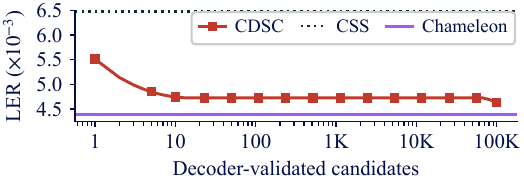}
\vspace{-10pt}
\caption{LER comparison in the binary-frame space for the $d$=$5$ surface code at $p=0.02$ under the Willow error map. CDSC cannot achieve optimum even after validating $10^5$ candidates with the decoder, whereas \design{} reaches the lowest LER with about $3\times10^3$ decoding-free search evaluations.}
\label{fig:randreach}
\end{figure}

We investigate how effective the CEM search is. We compare \design{} to decoder-validated CDSC with up to $10^5$ binary-frame candidates on the distance-5 surface code. We use $p{=}0.02$ to make decoder validation computationally tractable by increasing the frequency of logical failure events. Figure~\ref{fig:randreach} shows that CDSC cannot match \design{} even with $10^5$ candidates evaluated by the decoder.
However, \design{} achieves the lowest only with $3{\times}10^3$ decoder-free evaluations. This corroborates that good frames must be searched for rather than decoder-validated.

\subsection{Case Study 1: Memory Experiment}
We conduct a memory experiment to measure the logical performance of \design{} under various code families. Table~\ref{tab:mainprotocol} shows the detailed results. We indicate the gain of \design{} over the best prior baseline, measuring worst-axis LER. Across calibration-derived device maps, \design{} reduces the worst-axis LER relative to the best prior baseline by an average of $13\%$, $7\%$, and $4\%$ for surface, color, and BB codes, respectively. The differences across code families follow from the support structures of their ambiguity sets, which we measure directly. For the surface code, the class-$X$ and class-$Z$ ambiguity operators are separated, so a frame can reduce the LER of one class without increasing that of the other. The color code is self-dual: the two classes occupy identical supports, so a binary Hadamard frame exchanges the two-axis problems rather than solving either. 
For the BB codes, the smaller gains over the best prior baseline do not indicate a lack of deformation headroom: relative to the undeformed CSS frame, \design{} reduces LER by $19$--$58\%$ on average.
For all three code families, the largest gains occur on Willow, the most strongly biased device, indicating that the benefit increases with the strength of the spatial noise bias.

\begin{table*}[t]
\centering
\caption{Main Type-A LER Results, per-axis LER$_X$/LER$_Z$ for each prior baseline and \design{}, mean over calibration-map seeds at the operating point $p$ shown, every frame held to ${\ge}500$ ambiguity events per axis. \emph{Gain} is calculated as $\textit{Gain}=(LER_{\text{best\_baseline}}/LER_{\text{\design{}}}-1)*100\%$. The value in the gain column indicates mean [min,\,max] (\textit{higher is better}). }\vspace{-8pt}
\label{tab:mainprotocol}
\renewcommand{\arraystretch}{1.0}
\setlength{\tabcolsep}{1.8pt}

\begin{adjustbox}{max width=\textwidth}
\small
\begin{tabular}{|c|l|c||cc|cc|cc|cc||cc||cc||c|}
\hline
\multirow{2}{*}{Code} & \multirow{2}{*}{Noise} & \multirow{2}{*}{$p$} & \multicolumn{2}{c|}{CSS} & \multicolumn{2}{c|}{$XZZX$} & \multicolumn{2}{c|}{$ZXXZ$} & \multicolumn{2}{c||}{Local rule} & \multicolumn{2}{c||}{CDSC-24cand} & \multicolumn{2}{c||}{\textbf{\design{}}} & \multirow{2}{*}{\shortstack{Gain over best\\ prior baseline}} \\
\cline{4-15}
 & & & LER$_X$ & LER$_Z$ & LER$_X$ & LER$_Z$ & LER$_X$ & LER$_Z$ & LER$_X$ & LER$_Z$ & LER$_X$ & LER$_Z$ & LER$_X$ & LER$_Z$ &  \\
\hline\hline
\multirow{4}{*}{\shortstack{surf $d3$\\ $[[9,1,3]]$}}
& Berlin & 0.005 & 9.3e-4 & 8.7e-4 & 8.6e-4 & 8.5e-4 & 8.3e-4 & 9.5e-4 & 8.6e-4 & 8.7e-4 & 9.0e-4 & 9.0e-4 & 8.0e-4 & 8.4e-4 & +5\,[+2,+10] \\
\cline{2-16}
& Miami & 0.005 & 8.3e-4 & 6.4e-4 & 7.8e-4 & 6.9e-4 & 7.7e-4 & 7.0e-4 & 7.7e-4 & 7.1e-4 & 5.8e-4 & 6.4e-4 & 7.0e-4 & 6.4e-4 & +13\,[-1,+23] \\
\cline{2-16}
& Willow & 0.005 & 1.4e-3 & 7.8e-4 & 1.2e-3 & 9.2e-4 & 1.1e-3 & 1.1e-3 & 1.2e-3 & 9.1e-4 & 1.0e-3 & 1.1e-3 & 9.5e-4 & 9.1e-4 & +16\,[+4,+25] \\
\cline{2-16}
& xyz ($\eta{=}10$) & 0.005 & 1.1e-3 & 1.1e-3 & 1.1e-3 & 1.1e-3 & 6.7e-4 & 1.1e-3 & 1.2e-3 & 8.6e-4 & 9.4e-4 & 9.2e-4 & 4.5e-4 & 5.3e-4 & +106\,[+78,+157] \\
\hline\hline
\multirow{4}{*}{\shortstack{surf $d5$\\ $[[25,1,5]]$}}
& Berlin & 0.005 & 1.5e-4 & 8.8e-5 & 1.2e-4 & 1.1e-4 & 1.2e-4 & 1.0e-4 & 1.2e-4 & 1.1e-4 & 1.1e-4 & 1.1e-4 & 1.0e-4 & 1.0e-4 & +12\,[+2,+23] \\
\cline{2-16}
& Miami & 0.005 & 1.1e-4 & 7.9e-5 & 1.0e-4 & 9.5e-5 & 9.5e-5 & 1.0e-4 & 1.0e-4 & 9.7e-5 & 9.7e-5 & 9.2e-5 & 9.3e-5 & 9.0e-5 & +5\,[+0,+10] \\
\cline{2-16}
& Willow & 0.005 & 2.2e-4 & 7.6e-5 & 1.4e-4 & 1.3e-4 & 1.3e-4 & 1.4e-4 & 1.4e-4 & 1.3e-4 & 1.3e-4 & 1.2e-4 & 1.1e-4 & 1.2e-4 & +21\,[+8,+31] \\
\cline{2-16}
& xyz ($\eta{=}10$) & 0.005 & 1.5e-4 & 1.7e-4 & 1.7e-4 & 1.6e-4 & 1.6e-4 & 1.9e-4 & 1.8e-4 & 1.4e-4 & 1.3e-4 & 1.3e-4 & 5.3e-5 & 5.0e-5 & +206\,[+161,+262] \\
\hline\hline
\multirow{4}{*}{\shortstack{surf $d7$\\ $[[49,1,7]]$}}
& Berlin & 0.01 & 1.1e-4 & 7.2e-5 & 8.8e-5 & 9.7e-5 & 9.1e-5 & 8.7e-5 & 8.7e-5 & 9.3e-5 & 8.3e-5 & 9.3e-5 & 8.3e-5 & 7.7e-5 & +9\,[+0,+22] \\
\cline{2-16}
& Miami & 0.01 & 2.1e-4 & 1.4e-4 & 1.6e-4 & 1.8e-4 & 1.7e-4 & 1.7e-4 & 1.6e-4 & 1.9e-4 & 1.6e-4 & 1.7e-4 & 1.5e-4 & 1.5e-4 & +19\,[+15,+26] \\
\cline{2-16}
& Willow & 0.01 & 5.3e-4 & 1.1e-4 & 2.5e-4 & 2.4e-4 & 2.5e-4 & 2.4e-4 & 2.7e-4 & 2.4e-4 & 2.3e-4 & 2.4e-4 & 2.0e-4 & 2.1e-4 & +21\,[+9,+33] \\
\cline{2-16}
& xyz ($\eta{=}10$) & 0.01 & 3.0e-4 & 2.7e-4 & 2.8e-4 & 2.8e-4 & 2.7e-4 & 3.0e-4 & 3.2e-4 & 2.4e-4 & 2.4e-4 & 2.4e-4 & 7.1e-5 & 6.8e-5 & +357\,[+276,+421] \\
\hline\hline
\multirow{4}{*}{\shortstack{color $d3$\\ $[[7,1,3]]$}}
& Berlin & 0.005 & 7.5e-4 & 6.0e-4 & 6.2e-4 & 7.0e-4 & 7.3e-4 & 6.1e-4 & 6.3e-4 & 7.3e-4 & 6.9e-4 & 8.1e-4 & 6.4e-4 & 6.6e-4 & +2\,[-3,+8] \\
\cline{2-16}
& Miami & 0.005 & 6.3e-4 & 5.0e-4 & 5.2e-4 & 5.9e-4 & 6.0e-4 & 5.2e-4 & 5.4e-4 & 6.0e-4 & 5.4e-4 & 5.4e-4 & 5.3e-4 & 5.6e-4 & +3\,[-1,+8] \\
\cline{2-16}
& Willow & 0.005 & 1.0e-3 & 5.4e-4 & 6.5e-4 & 8.5e-4 & 8.9e-4 & 6.7e-4 & 6.8e-4 & 8.5e-4 & 6.6e-4 & 6.8e-4 & 7.0e-4 & 6.9e-4 & +20\,[-2,+53] \\
\cline{2-16}
& xyz ($\eta{=}10$) & 0.005 & 7.1e-4 & 5.8e-4 & 6.3e-4 & 5.7e-4 & 5.3e-4 & 6.3e-4 & 3.4e-4 & 8.7e-4 & 6.1e-4 & 5.8e-4 & 4.6e-4 & 6.0e-4 & +7\,[-5,+28] \\
\hline\hline
\multirow{4}{*}{\shortstack{color $d5$\\ $[[19,1,5]]$}}
& Berlin & 0.005 & 5.0e-5 & 3.8e-5 & 4.4e-5 & 4.4e-5 & 4.5e-5 & 4.4e-5 & 4.1e-5 & 4.4e-5 & 4.3e-5 & 4.4e-5 & 4.5e-5 & 4.6e-5 & -3\,[-8,+5] \\
\cline{2-16}
& Miami & 0.005 & 4.7e-5 & 3.3e-5 & 3.9e-5 & 4.2e-5 & 4.2e-5 & 3.9e-5 & 4.1e-5 & 4.2e-5 & 4.1e-5 & 4.1e-5 & 3.9e-5 & 4.2e-5 & +0\,[-3,+4] \\
\cline{2-16}
& Willow & 0.005 & 7.0e-5 & 2.2e-5 & 3.4e-5 & 4.5e-5 & 4.4e-5 & 3.3e-5 & 3.6e-5 & 4.0e-5 & 3.3e-5 & 4.3e-5 & 3.7e-5 & 3.8e-5 & +11\,[-2,+29] \\
\cline{2-16}
& xyz ($\eta{=}10$) & 0.005 & 2.9e-5 & 3.4e-5 & 4.1e-5 & 2.6e-5 & 2.6e-5 & 4.2e-5 & 2.7e-5 & 3.7e-5 & 3.4e-5 & 3.1e-5 & 2.9e-5 & 3.1e-5 & +10\,[-7,+20] \\
\hline\hline
\multirow{4}{*}{\shortstack{BB18\\ $[[18,4,4]]$}}
& Berlin & 0.01 & 4.8e-3 & 3.6e-3 & 4.3e-3 & 4.1e-3 & 4.2e-3 & 4.2e-3 & 4.3e-3 & 4.0e-3 & 4.5e-3 & 3.8e-3 & 4.2e-3 & 4.0e-3 & +1\,[-6,+6] \\
\cline{2-16}
& Miami & 0.01 & 4.0e-3 & 3.0e-3 & 3.5e-3 & 3.3e-3 & 3.4e-3 & 3.6e-3 & 3.6e-3 & 3.3e-3 & 3.5e-3 & 3.3e-3 & 3.5e-3 & 3.4e-3 & -2\,[-8,+10] \\
\cline{2-16}
& Willow & 0.01 & 6.0e-3 & 2.9e-3 & 4.3e-3 & 3.8e-3 & 3.9e-3 & 4.3e-3 & 4.2e-3 & 3.9e-3 & 4.5e-3 & 3.8e-3 & 4.0e-3 & 4.1e-3 & +0\,[-3,+3] \\
\cline{2-16}
& xyz ($\eta{=}10$) & 0.01 & 4.3e-3 & 4.6e-3 & 5.6e-3 & 4.1e-3 & 4.1e-3 & 5.4e-3 & 3.3e-3 & 4.3e-3 & 4.2e-3 & 4.0e-3 & 2.8e-3 & 2.7e-3 & +53\,[+27,+73] \\
\hline\hline
\multirow{4}{*}{\shortstack{BB36\\ $[[36,4,6]]$}}
& Berlin & 0.01 & 9.7e-5 & 5.5e-5 & 8.0e-5 & 6.7e-5 & 7.1e-5 & 7.7e-5 & 7.4e-5 & 6.8e-5 & 8.2e-5 & 6.5e-5 & 7.4e-5 & 7.3e-5 & +1\,[-3,+4] \\
\cline{2-16}
& Miami & 0.01 & 1.6e-4 & 8.6e-5 & 1.3e-4 & 1.1e-4 & 1.1e-4 & 1.2e-4 & 1.2e-4 & 1.1e-4 & 1.2e-4 & 1.1e-4 & 1.2e-4 & 1.1e-4 & -1\,[-6,+3] \\
\cline{2-16}
& Willow & 0.01 & 3.5e-4 & 8.8e-5 & 1.9e-4 & 1.6e-4 & 1.6e-4 & 1.8e-4 & 1.7e-4 & 1.6e-4 & 1.8e-4 & 1.6e-4 & 1.7e-4 & 1.7e-4 & -3\,[-20,+9] \\
\cline{2-16}
& xyz ($\eta{=}10$) & 0.01 & 2.0e-4 & 1.5e-4 & 1.8e-4 & 1.3e-4 & 1.4e-4 & 1.9e-4 & 1.3e-4 & 1.3e-4 & 1.5e-4 & 1.5e-4 & 1.2e-4 & 1.2e-4 & +26\,[-1,+64] \\
\hline\hline
\multirow{4}{*}{\shortstack{BB72\\ $[[72,12,6]]$}}
& Berlin & 0.01 & 8.4e-5 & 4.5e-5 & 6.8e-5 & 5.7e-5 & 6.5e-5 & 6.7e-5 & 7.2e-5 & 5.6e-5 & 7.1e-5 & 5.4e-5 & 6.6e-5 & 5.1e-5 & +3\,[-17,+29] \\
\cline{2-16}
& Miami & 0.01 & 1.1e-4 & 5.8e-5 & 8.4e-5 & 8.0e-5 & 8.5e-5 & 7.2e-5 & 8.4e-5 & 8.3e-5 & 9.3e-5 & 7.1e-5 & 7.9e-5 & 7.3e-5 & +3\,[-1,+6] \\
\cline{2-16}
& Willow & 0.01 & 3.3e-4 & 6.7e-5 & 1.8e-4 & 1.5e-4 & 1.8e-4 & 1.7e-4 & 2.0e-4 & 1.7e-4 & 2.0e-4 & 1.0e-4 & 1.4e-4 & 1.2e-4 & +32\,[+19,+51] \\
\cline{2-16}
& xyz ($\eta{=}10$) & 0.01 & 1.9e-4 & 1.4e-4 & 1.5e-4 & 1.6e-4 & 1.8e-4 & 1.4e-4 & 3.5e-4 & 3.8e-4 & 1.9e-4 & 1.2e-4 & 9.4e-5 & 8.9e-5 & +87\,[+39,+179] \\
\hline
\end{tabular}
\end{adjustbox}
\vspace{-13pt}
\end{table*}

\subsection{Case Study 2: Magic State Preparation}\label{sec:magic}
The worst-axis objective used elsewhere in this paper, $\max(\mathrm{LER}_X,\mathrm{LER}_Z)$, is appropriate for a stored logical qubit, which may be read in either basis. For the magic-state preparation considered here, only faults along one logical axis contribute to downstream failure~\cite{litinski2019}, so a frame chosen to balance both axes can sacrifice performance on the axis that matters. To evaluate single-axis frame optimization, we measure $\mathrm{LER}_Z$ on Willow at $p{=}0.005$ relative to $\textstyle{\mathrm{LER}_Z}$ with two-axis optimization. Table~\ref{tab:magic} shows a mean relative $\mathrm{LER}_Z$ of $0.34\times$--$0.67\times$ across the three code families relative to two-axis optimization. Thus, \design{} can select frames according to the logical error axis relevant to each workload.

\begin{table}[h]
\centering
\caption{Relative $\mathrm{LER}_{Z}$ under one-axis optimization with respect to two-axis optimization (mean [min,\,max]).}
\label{tab:magic}
\renewcommand{\arraystretch}{1.05}
\setlength{\tabcolsep}{4pt}
\small
\begin{tabular}{|l||c|c|c|}
\hline
Code & Mean & Range & \#noise maps \\
\hline\hline
Surface $d{=}5$ & $0.56\times$ & [$0.42\times$, $0.70\times$] & 10 \\
\hline
Color $d{=}5$ & $0.34\times$ & [$0.14\times$, $0.85\times$] & 10 \\
\hline
BB72 & $0.67\times$ & [$0.58\times$, $0.82\times$] & 10 \\
\hline
\end{tabular}
\end{table}

\subsection{Case Study 3: \design{} Mapped onto Devices}
\begin{figure}[htpb]
\centering
\vspace{-12pt}
\includegraphics[width=\columnwidth]{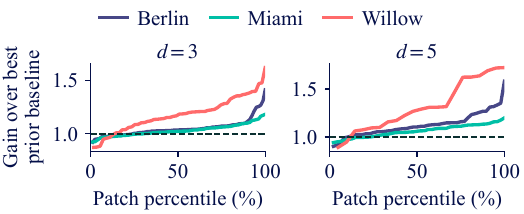}
\vspace{-20pt}
\caption{Across 332 valid surface code placements on the calibration maps of three devices under phenomenological noise, \design{} outperforms most patches, and the gain increases when the qubits in the patch are more biased.}
\label{fig:patchd3}
\end{figure}

We investigate whether the gain of \design{} is consistent with the physical placement constraints. We enumerate 332 admissible $d{=}3$ and $d{=}5$ surface-code patches on Berlin, Miami, and Willow, preserving device connectivity and evaluating each. 
Figure~\ref{fig:patchd3} shows that \design{} reduces worst-axis LER on 79\% of placements, with median and mean gains of 5.2\% and 7.6\%. Gains increase with distance on every device and are largest on Willow, where bias is strongest. Losses occur mainly on nearly uniform patches.

\subsection{Search-Algorithm Ablation }\label{sec:margin}
\begin{figure}[htpb]
\centering\includegraphics[width=\columnwidth]{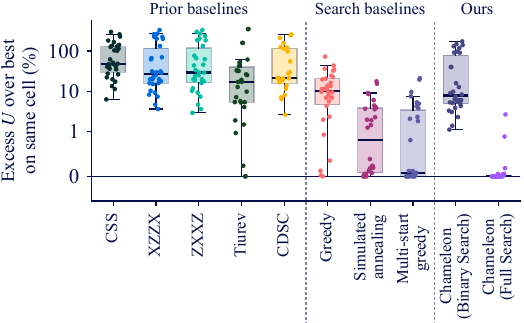}
\vspace{-15pt}
\caption{Impact of the cross-entropy method across the prior baselines and the search baselines, all at an equal evaluation budget (lower is better). \design{} consistently finds the lowest-$U$ frame.}
\label{fig:cem}
\vspace{-10pt}
\end{figure}

We evaluate whether \design{}'s performance depends on the choice of optimizer by comparing CEM against several search baselines under an equal surrogate-evaluation budget. The benchmark spans surface $d{=}3,5$, color $d{=}5$, BB18, and BB72 on Berlin, Willow, and the synthetic $\eta{=}10$ field at $p{=}0.005$.
Figure~\ref{fig:cem} shows \design{} against the prior baselines and three additional search baselines, all sharing the objective of Equation~\eqref{eq:memory-objective}. \design{} reduces the surrogate $U$ by $21$--$38\%$ and $2$--$12\%$ on average against the prior baselines and the search baselines, respectively. \design{} with the full frame search attains the lowest $U$.

\subsection{When Is Full-Frame Refinement Useful?}
\begin{figure}[h]
\centering\vspace{-10pt}
\includegraphics[width=\columnwidth]{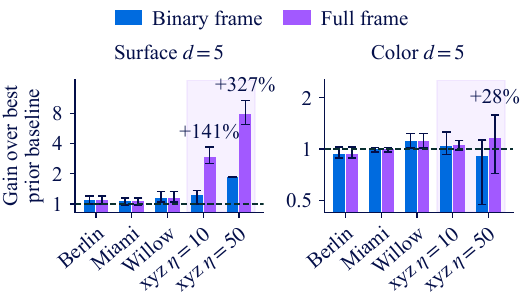}
\vspace{-15pt}
\caption{Binary frame search vs.\ full frame search at a physical rate of $p$=$0.005$. Full frame search is effective when $Y$-dominant qubits exist.}
\label{fig:sixframe}
\end{figure}

We evaluate when refinement from the binary frame space to the full frame space provides additional benefit. Figure~\ref{fig:sixframe} shows experiments conducted on $d$=5 surface and color codes.
For the current transmon-based devices, IBM Berlin, IBM Miami, and Google Willow, there are no $Y$-biased qubits. Thus, a full-frame search has no headroom to yield performance gains on these devices.
However, when $Y$-dominant qubits are present on a device, a full frame search becomes effective. It yields a small LER gain on moderately biased devices and a gain of up to $327\%$ on strongly biased devices.

\subsection{Sensitivity Study 1: Bias Strength and Physical Rate}\label{sec:res-sweeps}
\begin{figure}[!htb]
\centering
\vspace{-10pt}
\includegraphics[width=\columnwidth]{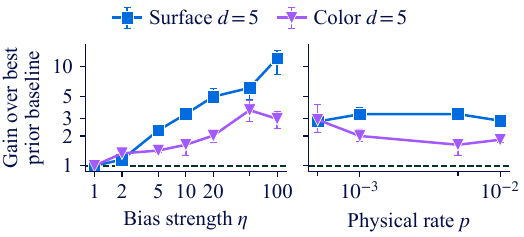}
\vspace{-20pt}
\caption{The gain holds across bias strength $\eta$ and rate $p$ on surface and color $d{=}5$. Mean over three maps. Bars span the 25th--75th percentile across maps. Both panels anchor at $\eta{=}10$/$p{=}0.005$.}
\label{fig:sweeps}
\vspace{-8pt}
\end{figure}

We study how sensitive \design{} is to bias strength and the physical rate.
Figure~\ref{fig:sweeps} shows the gain persists across bias strength and physical rate. The bias sweep varies $\eta$ on the xyz field. If the noise is unbiased ($\eta=1$), there is no LER gain, because a frame can help only when the noise is biased. 
The gain rises with bias to $9.5\times$ lower LER on surface and $2.9\times$ on color at $\eta{=}100$. In addition, \design{} consistently outperforms the baselines across a sweep of physical rate $p$.

\subsection{Sensitivity Study 2: Weight Cutoff in Enumeration}
\begin{figure}[htpb]
\centering
\vspace{-8pt}
\includegraphics[width=\columnwidth]{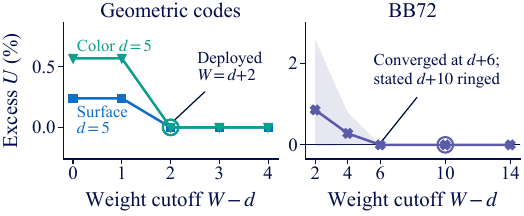}
\vspace{-20pt}
\caption{Sensitivity to the weight cutoff $W$ (lower $U$ is better).}
\label{fig:enum_sens}
\vspace{-10pt}
\end{figure}

We investigate how deep the ambiguity enumeration must run. Figure~\ref{fig:enum_sens} sweeps the weight cutoff $W$, scoring each cutoff by the frame it selects. All geometric codes converge at the deployed $W{=}d{+}2$, and BB72 converges at $W{=}d{+}6$, so $W{=}d{+}10$ is a safety margin for the cutoff.

\subsection{Sensitivity Study 3: Deployment Margin}
\begin{figure}[htpb]
\centering
\vspace{-12pt}
\includegraphics[width=\columnwidth]{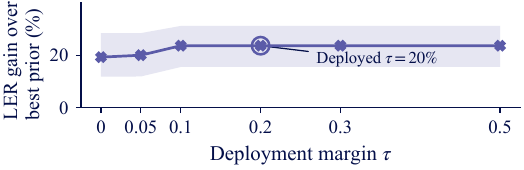}
\vspace{-23pt}
\caption{Impact of the deployment margin $\tau$ on BB72.}
\label{fig:sens_margin}
\end{figure}

We evaluate how sensitive the deployment margin $\tau$ is. Figure~\ref{fig:sens_margin} sweeps $\tau$ and the corresponding LER on BB72 and Willow. Always deploying the full-frame candidate ($\tau{\le}5\%$) yields $+12.0\%$ over the baselines. With $\tau{\ge}10\%$, \design{} chooses the binary frame and holds $+23.5\%$ gain.
\section{Discussion}
\subsection{\design Under Circuit-Level Noise}

We conduct experiments using the phenomenological error model to evaluate logical performance because it is computationally efficient and allows us to isolate the performance of frame selection.
To determine whether \design works under circuit-level noise, we adopt the \texttt{stim::noise::Si1000} noise model from Stim~\cite{gidney2021stim}. We change only the physical error rate to $p$=$0.005$ and use a biased idle-error model derived from Google Willow calibration data.

\begin{figure}[htpb]
\centering
\includegraphics[width=\columnwidth]{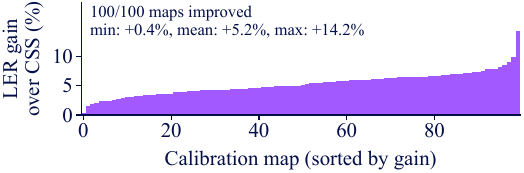}
\vspace{-15pt}
\caption{LER gain of \design over CSS for a distance-5 surface code across 100 valid calibration maps sampled from Google Willow calibration data. \design consistently records lower LER (95\% CI: 0.4\%).}
\label{fig:circuit}
\end{figure}

Figure~\ref{fig:circuit} shows that \design consistently reduces LER by 5.2\% relative to CSS on average and by up to 14\%. However, the gain is smaller than that under the phenomenological model. Because the idle error is the only biased component, while the other components follow uniform noise, the overall level of bias is weak.
This result may change under a more precise biased-noise model. However, evaluating such a model requires access to actual hardware, and we do not have access to this superconducting hardware. We leave this direction for future work.

\subsection{Complexity Analysis of \design}\label{sec:complexity}

\design{} constructs its LUT once per code offline and reuses it for every online, per-map frame search. The offline cost, including combinatorial enumeration, is therefore amortized across maps. During online search, the LUT is read-only: scoring one frame costs $O(|\mathcal{L}|n)$, and a CEM search with $K$ candidates costs $O(K|\mathcal{L}|n)$. The online stage samples the $6^n$ frame space rather than enumerating it. Table~\ref{tab:complexity} reports each stage’s time, memory, and payment schedule. Here, $T_{\text{GE}}(m,n)$ denotes the time cost of one randomized Gaussian elimination.
\begin{table}[htpb]
\centering
\caption{Time and memory of the \design{} pipeline with its \emph{payment schedule} ($n$ qubits, $m$ checks, $|\mathcal{L}|$ ambiguity operators, $K$ CEM candidates, $R$ BB72 trials, and $s$ batch size).}
\label{tab:complexity}

\renewcommand{\arraystretch}{1.01}
\setlength{\tabcolsep}{1pt}

\begin{adjustbox}{max width=\columnwidth}
\small
\begin{tabular}{|l||c|c|c|c|}
\hline
Stage & Paid & Time & Measured & Memory \\
\hline\hline

\makecell[l]{LUT build\\(geometric)}
&
\makecell{Offline\\once/code}
&
$O\bigl(n^W\bigr)$
&
\makecell{$8$\,min (surf $d7$)\\$2.3$\,min (color $d7$)}
&
\makecell{LUT $O(|\mathcal{L}|\,n)$\\a few MiB}
\\
\hline

\makecell[l]{LUT build\\(BB72)}
&
\makecell{Offline\\once/code}
&
\makecell{$O(R\cdot T_{\rm GE}(m,n)$\\$+|\mathcal{L}|n)$}
&
$<30$\,s
&
\makecell{LUT $O(|\mathcal{L}|\,n)$\\tens of MiB}
\\
\hline

\makecell[l]{Score one\\frame by $U$}
&
\makecell{Online\\per cand.}
&
$O(|\mathcal{L}|\,n)$
&
\makecell{$\le 2.5$\,ms (BB72)\\$\mu$s (geometric)}
&
\makecell{shared LUT\\$+O(n)$}
\\
\hline

\makecell[l]{CEM\\search}
&
\makecell{Online\\per map}
&
$O(K|\mathcal{L}|\,n)$
&
\makecell{$K=75,000$:\\$<1$\,s--$3.1$\,min}
&
\makecell{shared LUT\\$+O(s\,n)$}
\\
\hline

\end{tabular}
\end{adjustbox}

\vspace{-6pt}
\end{table}
\newpage
\section{Related Work}\label{sec:related}

\noindent \textbf{Bias-Tailored Codes.} Biased-noise fault tolerance fixes one global frame~\cite{aliferis2009biased,aliferis2008fault}, spanning tailored XY~\cite{tuckett2018ultrahigh,tuckett2019tailoring,tuckett2020thresholds}, XZZX/ZXXZ~\cite{bonillaataides2021xzzx}, bias-tailored LDPC~\cite{roffe2023biastailored}, rectangular patches~\cite{azad2022}, domain-wall color codes~\cite{tiurev2024domainwall}, cat-qubit XZZX~\cite{darmawan2021} under uniform bias and as a code design rather than a per-qubit objective~\cite{xu2023tailoredxzzx}. 
\design{} optimizes per qubit against the measured map instead.

\vspace{0.05in}
\noindent \textbf{Per-Qubit Deformation.} Per-qubit deformation otherwise stops short of a coordinated search against the measured map. CDSC samples random deformations and decodes each to keep the best~\cite{dua2024clifford}, a per-map decoder budget that is infeasible at scale. The Tiurev rule assigns each qubit greedily from its local bias rather than jointly~\cite{tiurev2023}, and compass-code patterns are hand-crafted~\cite{campos2026compass}. Zero-rate qLDPC deformation certifies uniform-bias thresholds as a proof device, not a search objective~\cite{das2026cliffordldpc}. Robertson et al.\ brute-force decode against a measured model~\cite{robertson2017tailored}, an ancestor of our mechanism sweep, and RL agents discover codes under uniform models~\cite{olle2024,nautrup2019}. None runs a coordinated search against a spatially non-uniform map, whereas \design{} addresses it.

\section{Concluding Remarks}\label{sec:concl}
We present a fast, high-performance, code-agnostic compiler for Clifford deformation that addresses biased noise.
To make compilation computationally feasible, we introduce a surrogate objective that replaces the ambiguity likelihood with an upper bound. We also precompute code-specific ambiguity LUTs from sets of ambiguity operators, enabling their reuse across compilation instances. By separating this offline computation from frame selection, our approach reduces frame-selection latency from days to seconds or minutes.
Across surface, color, and BB codes, \design{} reduces the worst-axis LER relative to the best prior baseline by an average of 13\%, 7\%, and 4\%, respectively, in memory experiments. Moreover, single-axis optimization provides an additional 50--191\% improvement. These gains increase as the noise bias becomes stronger, persist as physical error rates decrease, and remain under circuit-level noise.

\vfill

\section*{\design{} Algorithm}\label{app:alg}
Algorithm~\ref{alg:cham} comprises one offline stage for code-static ambiguity-set construction and two online stages: binary-frame search with the analytic bound and full-frame refinement.
\begin{algorithm}[b]
\small
\caption{\design: Calibration-aware frame selection}\label{alg:cham}
\begin{algorithmic}[1]
\State \textbf{Input}: code $(H_X,H_Z,\{\bL_{X,i}\},\{\bL_{Z,i}\})$, map~$(p_X,p_Y,p_Z)_q$
\State \textbf{Output}: optimized frame $F^\star$
\Statex {\color{gray}\small //\;\S\ref{sec:mechs} Offline: code-static ambiguity-set construction}
\State $\mathcal{A}_X,\mathcal{A}_Z\gets\mathrm{LowWtLogicals}(H_X,H_Z,\{\bL_{X,i}\},\{\bL_{Z,i}\})$
\Statex {\color{gray}\small //\;\S\ref{sec:impact-surrogate},\S\ref{sec:search} Online: per-map frame search}
\State $U(F)\gets\max_{c\in\{X,Z\}}\sum_{\ell\in\mathcal{A}_c}\Gamma_\ell(F)$ \textbf{or} $\sum_{\ell\in\mathcal{A}_{c^\star}}\Gamma_\ell(F)$
\State $F_2^\star\gets\mathrm{CEM}\bigl(U,\{I,H\}^n\bigr)$
\State $\widehat{F}_6\gets\mathrm{CEM}\bigl(U,S_3^n;F_2^\star\bigr)$
\State $F^\star\gets\widehat{F}_6$ if $U(\widehat{F}_6)\le(1-\tau)U(F_2^\star)$, else $F_2^\star$; deploy $F^\star$
\end{algorithmic}
\end{algorithm}

\section*{Acknowledgements}
Won Joon Yun appreciates the valuable discussion with Sayam Sethi, Maxwell Poster, and Aditi Awasthi and the editorial feedback from Sarvesh Raghuraman. 

\section*{Derivation of Analytical Bound}  We provide the full derivation of the analytic bound used in Equation~\eqref{eq:operator-bound}. Consider a logical class $c\in\{X,Z\}$, a frame $F$, and an ambiguity operator $\ell$. For each logical class, we represent an error pattern by a binary vector $e\in\{0,1\}^n$, where $e_q=1$ denotes a class-$c$ error on qubit $q$. The error pattern competing pairwise with $e$ is represented by the ambiguity operator, \textit{i.e.}, $e' = e\oplus\ell$.
The two patterns have the same syndrome but different logical effects. A logical failure associated with the operator $\ell$ occurs when the competing pattern $e'$ is at least as likely as the true pattern $e$. Thus,  
\begin{equation} 
\textstyle{\Pr_\ell[\text{logical failure}] = \sum_e \Pr(e)\cdot \mathbf{1}[\Pr(e\oplus\ell)\geq\Pr(e)]. }    \label{eq:logical_failure} 
\end{equation}  First, we bound each term individually. Whenever the indicator $\mathbf{1}(\cdot)$ is one, $\Pr(e\oplus\ell)\geq\Pr(e)$, and therefore $\Pr(e)\leq\sqrt{\Pr(e)\Pr(e\oplus\ell)}$. Thus, the probability of logical failure associated with the operator $\ell$ is bounded as,   
\begin{equation}    
\textstyle{\Pr_\ell[\text{logical failure}] \leq \sum_e \sqrt{\Pr(e)\Pr(e\oplus\ell)}}.   \label{eq:logical_failure_bound} 
\end{equation}  
Under independent per-qubit noise, the probability of an error pattern factors across qubits. Therefore, the RHS of Equation~\eqref{eq:logical_failure_bound} becomes  \begin{equation} \sum_e \sqrt{\Pr(e)\Pr(e\oplus\ell)} = \prod_q \sum_{e_q\in\{0,1\}} \sqrt{\Pr(e_q)\Pr(e_q\oplus\ell_q)}.  \label{eq:logical_failure_per_qubit_noise} \end{equation}  

Consider each contribution from each qubit. If $q\notin\ell$, then $\ell_q=0$, and the term in RHS of Equation~\eqref{eq:logical_failure_per_qubit_noise} is written as
\begin{equation} 
\sum_{e_q\in\{0,1\}} \sqrt{\Pr(e_q)\Pr(e_q)} = \sum_{e_q\in\{0,1\}}\Pr(e_q)=1. 
\end{equation}  
If $q\in\ell$, then $\ell_q=1$. 
Let $r=r_c^q(F)$, the contribution from each qubit is 
\begin{equation} 
\sqrt{(1-r)r}+\sqrt{r(1-r)} = 2\sqrt{r(1-r)} = \gamma(r). 
\end{equation}  
Therefore, all qubits outside the support of $\ell$ contribute a factor of $1$, while each qubit in the support of $\ell$ contributes $\gamma(r_c^q(F))$. 
Thus,  $\Pr_\ell[\text{logical failure}] \leq \prod_{q\in\ell} \gamma\!\left(r_c^q(F)\right),$ which gives Equation~\eqref{eq:operator-bound}.
 
\bibliographystyle{ieeetr}
\bibliography{refs}
\end{document}